\documentclass[aps,pra,reprint,superscriptaddress]{revtex4-2}

\usepackage[final]{changes}

\usepackage{siunitx}
\usepackage{graphicx}
\usepackage{braket}
\usepackage{hyperref}

\begin{document}


\title{Reservoir-based deterministic loading of single-atom tweezer arrays}


\author{Lars Pause}
\affiliation{Technische Universit\"at Darmstadt, Institut f\"ur Angewandte Physik, Schlossgartenstra\ss e 7, 64289 Darmstadt, Germany}
\author{Tilman Preuschoff}
\affiliation{Technische Universit\"at Darmstadt, Institut f\"ur Angewandte Physik, Schlossgartenstra\ss e 7, 64289 Darmstadt, Germany}
\author{Dominik Sch\"affner}
\affiliation{Technische Universit\"at Darmstadt, Institut f\"ur Angewandte Physik, Schlossgartenstra\ss e 7, 64289 Darmstadt, Germany}
\author{Malte Schlosser}
\affiliation{Technische Universit\"at Darmstadt, Institut f\"ur Angewandte Physik, Schlossgartenstra\ss e 7, 64289 Darmstadt, Germany}
\author{Gerhard Birkl}
\email[]{Contact for correspondence: apqpub@physik.tu-darmstadt.de}
\homepage[]{https://www.iap.tu-darmstadt.de/apq}
\affiliation{Technische Universit\"at Darmstadt, Institut f\"ur Angewandte Physik, Schlossgartenstra\ss e 7, 64289 Darmstadt, Germany}
\affiliation{Helmholtz Forschungsakademie Hessen f\"ur FAIR (HFHF), Campus Darmstadt, Schlossgartenstra\ss e 2,
	64289 Darmstadt, Germany}
\date{\today}

\begin{abstract}
State-of-the-art individual-atom tweezer platforms have relied on loading schemes based on spatially superimposing the tweezer array with a cloud of cold atoms created beforehand. Together with immanent atom loss, this dramatically limits the data rate, as the application sequence must be alternated with the time-consuming phases of magneto-optical trapping and laser cooling. We introduce a modular scheme built on an additional cold-atom reservoir and an array of buffer traps effectively decoupling cold-atom accumulation and single-atom supply from the quantum-register operation. For this purpose, we connect a microlens-based tweezer array to a cloud of laser-cooled atoms held in an auxiliary large-focus dipole trap by utilizing atom transport and buffer traps for dedicated single-atom supply. We demonstrate deterministic loading of a hexagonal target structure with atoms
solely originating from the reservoir trap.
The results facilitate increased data rates and unlock a path to continuous operation of individual-atom tweezer arrays in quantum science, 
making use of discrete functional modules, operated in parallel and spatially separated.\\
\textbf{Published as \doi{10.1103/PhysRevResearch.5.L032009}} 
\end{abstract}


\maketitle

\begin{figure}
\includegraphics[width=0.45 \textwidth]{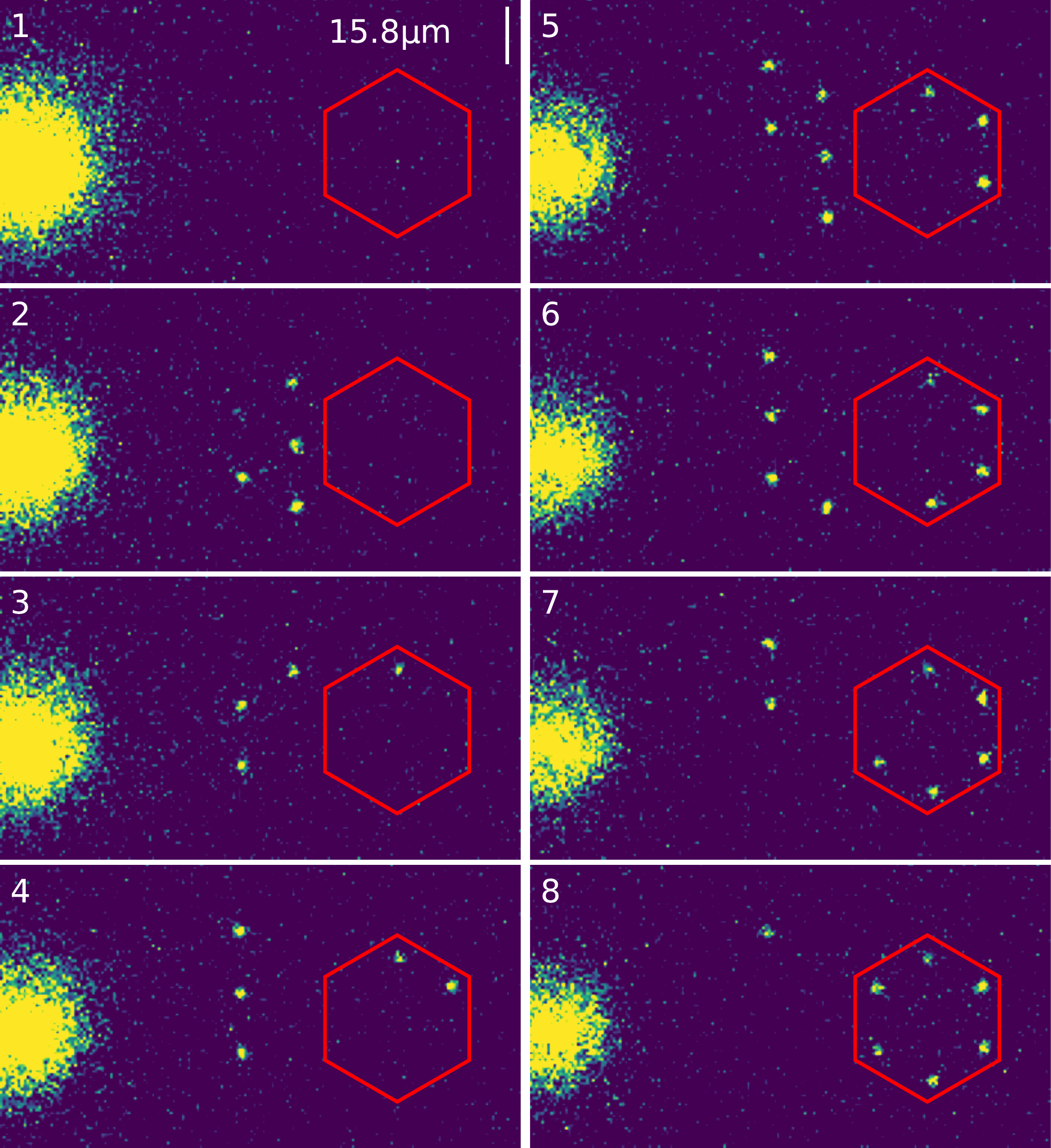}
\caption{\label{fig:atom_by_atom} Series of consecutive fluorescence images of a single experimental realization when loading one atom at a time into a hexagonal target structure (red): Frame 1 shows the filled reservoir trap and the empty tweezer array. In frame 2, the buffer traps have been stochastically loaded with individual atoms from the reservoir. The subsequent frames show the deterministic atom-by-atom loading of the target hexagon.}
\end{figure}

\begin{figure}[t]
\includegraphics[width=0.48 \textwidth]{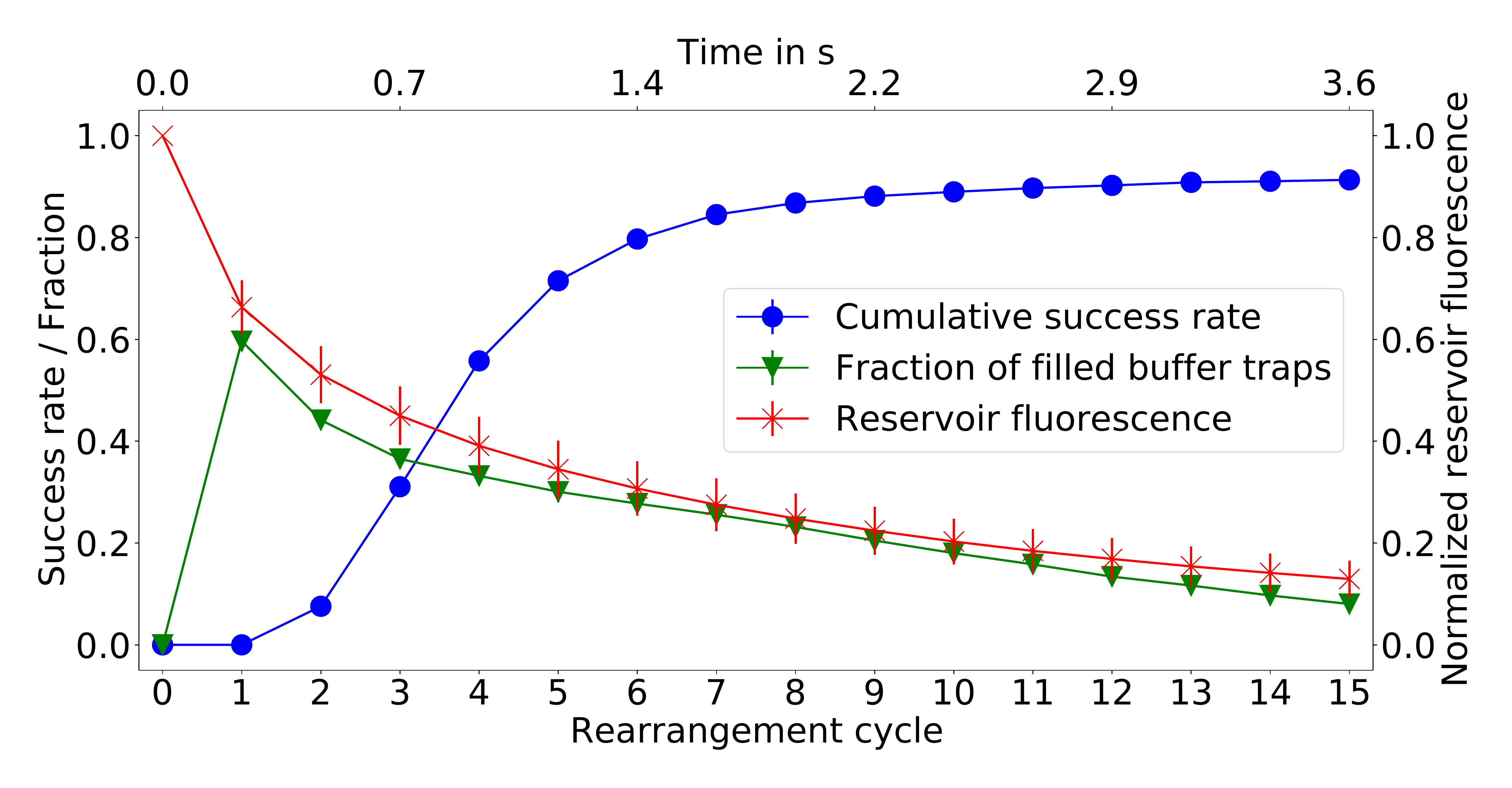}
\caption{\label{fig:atom_no_evolution_20220225} Quantitative analysis of the reservoir and buffer-trap-based loading of tweezer arrays as a function of the number of rearrangement cycles. Data are based on $>$2500 experimental repetitions: (Blue) Cumulative success rate of a defect-free target structure of six sites in a hexagonal arrangement. (Red) Normalized reservoir signal corresponding to the atom number inside the reservoir trap. (Green) Average fraction of filled buffer traps available to refill empty target traps. Empty buffer traps are refilled from the reservoir trap during each rearrangement cycle. For the normalized reservoir signal, the uncertainty is given by the standard deviation and for the two other signals by the $1\sigma$ confidence interval (error bars might be smaller than the size of the data points).}
\end{figure}

Quantum devices based on individual atoms trapped in large-scale registers of optical potentials have shown their high potential as  platforms for quantum simulation and quantum computation in recent years \cite{Adams2019,Browaeys2020,Henriet2020,Morgado2021,Kaufman2021,Shi2022,Cong2022,Wintersperger2023}. Based on spatial light modulators, multitone acousto-optic deflectors (AODs), or as in our approach, microlens arrays (MLAs), \added{defect-free structures of neutral atoms have been prepared in one dimension \cite{Endres2016a,Levine2019,Madjarov2020,Ma2022}, two dimensions (2D) \cite{Dumke2002,Barredo2016,Brown2019,OhldeMello2019,Wang2020,Kim2020,Ebadi2021,Scholl2021,Sheng2022,Graham2022,Barnes2022,Schymik2022,Yan2022,Tian2023}, and three dimensions \cite{Barredo2018,Kumar2018,Schlosser2023}}. The scalability of these implementations has been demonstrated up to a few hundred qubits \cite{OhldeMello2019,Ebadi2021,Scholl2021,Schymik2022,Schlosser2023}, \added{while recent advancements have opened a route to controlling thousands of single-atom quantum systems \cite{Huft2022,Schlosser2023}}. Furthermore, utilizing Rydberg interactions, the application of 2-qubit operations as well as the simulation of spin Hamiltonians is possible with neutral-atom quantum systems \cite{Levine2019,Madjarov2020,Kim2020,Ebadi2021,Scholl2021,Graham2022,Ma2022,Chew2022,McDonnell2022,Steinert2022}.\\
\indent
One remaining challenge of neutral-atom platforms is to access high data rates. A typical experimental cycle starts by preparing a cold-atom ensemble which is loaded into a 2D configuration of dipole potentials. Utilizing collisional blockade, 60\% filling with individual atoms is typical \cite{Schlosser2002}. By implementation of gray molasses or dark-state enhanced loading, filling fractions of up to 90\% in a system with 100 trapping sites have been demonstrated \cite{Brown2019,Jenkins2022,Shaw2023}. Nevertheless, the intrinsic stochastic component limits the performance. Thus, after the loading process, atom resorting is necessary to achieve defect-free structures \cite{Schymik2020,Sheng2021,Cimring2022,Sabeh2022,Tian2023}.  Each time a prepared structure develops a defect, resorting has to be repeated, and in the case of insufficient spare atoms, the experimental cycle must be started again with cold-atom preparation. Moreover, the duty cycle is further reduced by qubit readout operations utilizing loss-based detection schemes demanding a new preparation of the target structure after each experimental cycle. This affirms an urgent demand for advanced concepts of individual atom supply.\\
\begin{figure}
\includegraphics[width = 0.45 \textwidth]{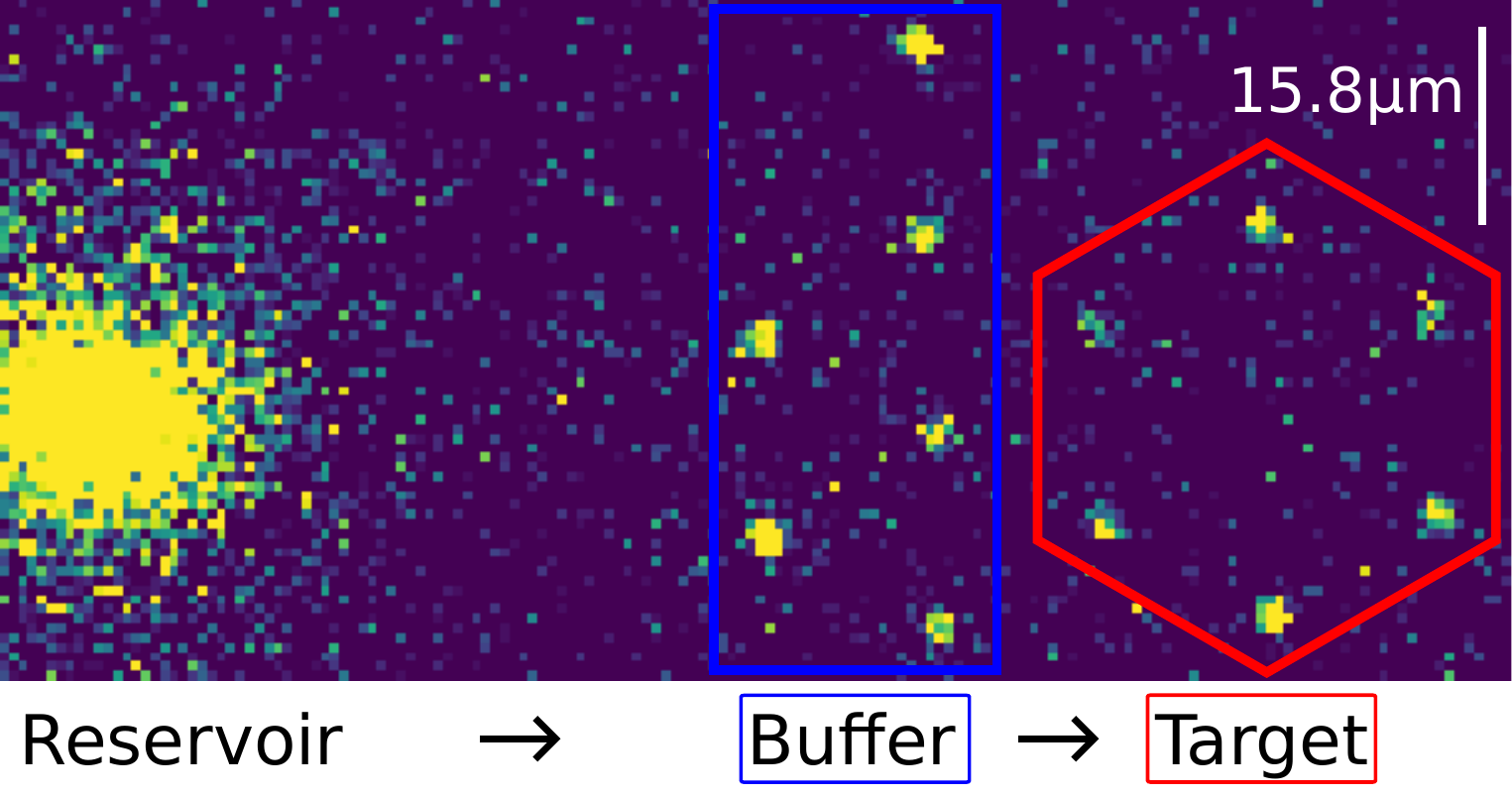}
\caption{\label{fig:Principle} Single-shot fluorescence image of individual atoms in the hexagonal tweezer array (right) and an atom ensemble in the reservoir dipole trap (left). The blue frame defines the buffer-trap section and the red frame the target structure. The image is taken after four rearrangement sequences and shows a fully filled target structure.}
\end{figure}%
\indent
In the literature, the preparation of registers of individual neutral atoms in large tweezer arrays is based on loading atoms directly from optical molasses being spatially superposed with the whole array. Enhanced loading of atom ensembles has been demonstrated by overlapping a pancake-shaped dipole trap for increased atom density \cite{Wang2020}. The spatial overlap of the cold-atom preparation stage and the tweezer array intrinsically limits the duty cycle as magneto-optical trapping (MOT) and molasses phases need to be alternated with single-atom preparation and application sequences.
\begin{figure*}[t]
\includegraphics[width = \textwidth]{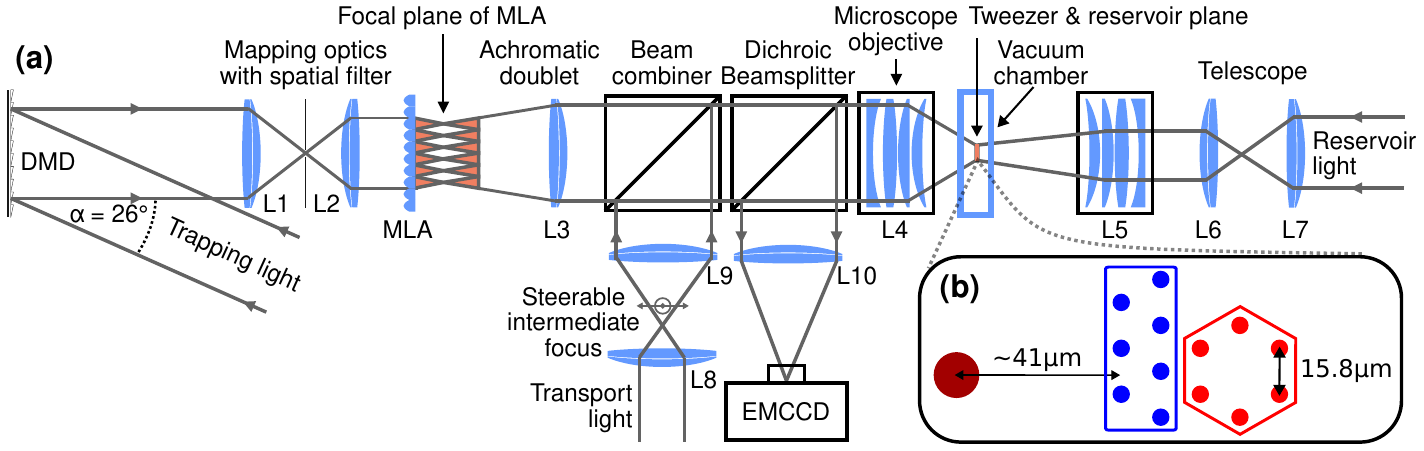}
\caption{\label{fig:exp_setup}Reservoir and hexagonal tweezer array setup (not to scale) for deterministic filling of single-atom tweezer arrays. (a) A digital micromirror device (DMD) is illuminated by the trapping light such that the third diffraction order is deflected orthogonally off the DMD surface. Using lenses L1 and L2 the DMD image is mapped onto the microlens array (MLA) whose focal plane is demagnified into the vacuum chamber using the relay optics L3 and L4. With a beam combiner, an acousto-optic deflector (AOD)-based movable laser beam acting as transport tweezers is overlapped with the trapping beam. The reservoir beam, counterpropagating to the tweezer-array laser beam, is focused in the tweezer plane by L5 (equivalent to L4). The waist of this single-beam dipole trap can by adjusted by telescope lenses L6 and L7. The fluorescence light of the trapped atoms is recorded by an electron-multiplying charge-coupled device (EMCCD) camera. Using the DMD, microlenses can be selectively addressed to achieve the trap pattern used in this letter. (b) A schematic of the resulting trap configuration in the tweezer and reservoir plane. The atom reservoir is shown in dark red, and the two tweezer array sections used as buffer and target traps are shown in blue and red, respectively.}
\end{figure*}
In this letter, we present the concept of separating the sequential steps for defect-free filling of optical tweezers with individual atoms into different functional modules (Fig.~\ref{fig:Principle}) designed to be operated sequentially or in parallel. This paves the way for a continuous supply of cold atoms prepared at different times \cite{Singh2022} and at different sections of the apparatus \cite{Schlosser2012}.\\
\indent
Here, the complete process of deterministic loading of dipole-trap target structures with individual atoms is separated into four modules: (1) preparation of ultracold atoms in a MOT and optical molasses, (2) transfer of a large cloud of ultracold atoms into a large focused-beam dipole trap, i.e., a reservoir trap, (3) extraction of a small ensemble of atoms using movable transport tweezers and application of collisional blockade during transport to a buffer trap section, and finally, (4) deterministic loading of the target structure with individual atoms from the buffer traps, again using transport tweezers. 
In Fig.~\ref{fig:Principle}, a single-shot fluorescence image of trapped atoms exemplifies the overall trap architecture used in this letter: A large focused-beam dipole trap (left) acts as a cold-atom reservoir from which atoms are extracted by the transport tweezers and positioned in the buffer-trap section (blue frame) of a hexagonal array of optical dipole traps. Individual atoms in the buffer traps are then utilized to deterministically fill the target structure consisting of a dipole trap hexagon (red frame).\\ 
\indent
The schematic of the experimental setup is presented in Fig.~\ref{fig:exp_setup}(a) with the tweezer and reservoir planes shown in Fig.~\ref{fig:exp_setup}(b) in detail. This plane is oriented normal to the central optical axis of the setup.
We prepare several $10^4$ laser-cooled $^{85}$Rb atoms in a custom-made vacuum chamber using six MOT beams and a deceleration beam, all with a $1/e^2$ radius of $\SI{2.5(1)}{\milli\metre}$. The atoms are supplied by a rubidium dispenser and slowed using the deceleration beam with an intensity of $6.7(6)~I_{\mathrm{sat}}$ and a fixed detuning of $-3.5~\Gamma$ ($\Gamma = 2\pi \times \SI{6.07}{\mega\hertz}$) relative to the $\Ket{5S_{1/2}, F=3} \leftrightarrow \Ket{5P_{3/2}, F=4}$ cooling transition. The MOT consists of  
four beams with an intensity of $22(2)~I_{\mathrm{sat}}$ per beam, forming a rectangular cross in a plane oriented at an angle of $\SI{15}{\degree}$ relative to the tweezer plane and two beams with $30(3)~I_{\mathrm{sat}}$ per beam, irradiating the atoms at an angle of $\SI{47}{\degree}$ relative to the tweezer plane. All MOT beams have detuning of $-2.0~\Gamma$. The MOT is loaded for \SI{1.8}{\second}, with the deceleration beam being turned off during the final \SI{300}{\milli\second}. The MOT phase is followed by a molasses phase of typically \SI{40}{\milli\second} in which the magnetic field is turned off, the cooling laser intensities are reduced to $\frac{1}{10}$ of the MOT intensities, and the detuning is changed to $-5.0~\Gamma$. This results in an atom cloud with a temperature of \SI{22(5)}{\micro\kelvin}.\\ 
\indent
A hexagonal array of optical tweezers is created by reimaging the focal plane of a hexagonal MLA with a \SI{750}{\milli\metre} achromatic doublet (L3) and a microscope objective (L4) with a focal length of \SI{37.5(10)}{\milli\metre} \added{and a numerical aperture (NA) of 0.25(2)} into the vacuum chamber. The fused silica MLA has a pitch of \SI{161.5}{\micro\metre} with a radius of curvature of \SI{2.05}{\milli\metre} for each lenslet. Reimaging creates foci with a pitch of \SI{7.9(1)}{\micro\metre} and a waist of \SI{2.0(2)}{\micro\metre} in the tweezer plane. For \added{selecting a subset of the hexagonal trap pattern and dynamically adjusting} the light field illuminating each lenslet, we place a digital micromirror device (DMD) illuminated by a Gaussian laser beam with wavelength of \SI{796.5}{\nano\metre} and $1/e^2$ radius of $\SI{1.2(1)}{\milli\metre}$ in front of the MLA~\cite{Schaeffner2020}. The DMD is mapped onto the MLA by two achromatic doublets with focal lengths of \SI{50}{\milli\metre} (L1) and \SI{30}{\milli\metre} (L2), respectively. 
With micromirrors of $7.64~\times$~\SI{7.64}{\micro\metre\squared} size, this results in typically 875 mirrors illuminating each lenslet.
For the experiments presented in this letter, we programed the DMD to double the pitch of the tweezer array to \SI{15.8(2)}{\micro\metre} by turning off the illumination of every second lenslet and to generate a configuration of seven buffer traps  and a hexagonal target structure of six traps, as shown in Figs.~\ref{fig:Principle} and \ref{fig:exp_setup}(b). The deepest trap has a trap depth of $U_{\mathrm{array}}/k_{\mathrm{B}} = \SI{600(200)}{\micro\kelvin}$. In addition, by adjusting the number of micromirrors illuminating each lenslet, the variation of trap depths within the array is reduced to $\sim$25\%, thus counteracting the intensity variation given by the Gaussian beam profile of the trapping light.\\ 
\indent
In the modular sequence introduced here, atoms are not loaded directly from the molasses into the trap array, as usual, but are transferred into a large-focus reservoir trap first by superimposing the reservoir potential with the optical molasses for \SI{20}{\milli\second}.
The reservoir trap is created by light at a wavelength of \SI{798.8}{\nano\metre} which is counterpropagating with respect to the trap array beam and focused in the tweezer plane using a second microscope objective (L5). This objective has the same parameters as L4 and is used to create a focus with $1/e^2$ waist of $\SI{14.6(1)}{\micro\metre}$ at a lateral distance of \SI{41}{\micro\metre} from the buffer trap array. The typical trap depth is $U_{\mathrm{reservoir}}/k_{\mathrm{B}} = \SI{600(200)}{\micro\kelvin}$. The reservoir occupation is determined by illuminating the atoms with cooling light red-detuned by $-5.0~\Gamma$ relative to the cooling transition and repumping light red-detuned by $-0.2~\Gamma$ relative to the $\Ket{5S_{1/2}, F=2} \leftrightarrow  \Ket{5P_{3/2}, F=3}$ repumping transition. The resulting flourescence light is recorded for \SI{100}{\milli\second} with an electron-multiplying charge-coupled device (EMCCD) camera using the objective L4 in combination with a \SI{750}{\milli\metre} achromatic doublet (L10). \added{The imaging time is limited by the NA of the objective L4 as well as the detection efficiency of the camera \cite{OhldeMello2019}.} The outcome of the resulting stage (2) of the modular sequence gives a reservoir trap typically filled with $\sim$80 atoms and no atoms in the microlens-based tweezer array [see Fig.~\ref{fig:atom_by_atom}~(Frame 1)]. \\ 
\indent
Using movable transport tweezers at a wavelength of \SI{796.5}{\nano\metre}, atoms can be transferred from the reservoir trap to the tweezer array or rearranged within the different segments of the array. 
The transport tweezers are created using a collimated beam with $1/e^2$ radius of $\SI{1.5(1)}{mm}$ propagating through a 2D AOD that deflects the beam in both lateral dimensions. The beam is focused by a \SI{60}{\milli\metre} achromatic doublet (L8) to translate the angular deflection into a  linear translation. The focus is reimaged into the vacuum chamber by propagating the beam through a \SI{400}{\milli\metre} achromatic doublet (L9) and overlapping it with the trap array beam before passing through the objective L4. This creates transport tweezers with $1/e^2$ waist of \SI{2.2(1)}{\micro\metre} and a scanning range of \SI{250}{\micro\metre} in both lateral dimensions within the tweezer plane.\\ 
\indent
We use atoms from the reservoir to load the tweezer array in the next two stages of our modular sequence. For this purpose, we divide the trap array into a region of buffer traps and a target trap section: The buffer traps, marked in blue in Figs.~\ref{fig:Principle} and \ref{fig:exp_setup}(b), are stochastically loaded from the reservoir and deterministically provide individual atoms for subsequently loading the target structure marked in red. 
For filling empty buffer traps from the reservoir, we overlap the transport tweezers with the center of the reservoir trap, ramp up the transport tweezer depth to $U_{\mathrm{T}}/k_{\mathrm{B}} = \SI{800(200)}{\micro\kelvin}$, and move the trapped atoms with the tweezers to an empty buffer trap position where they are released by ramping down the transport tweezers to zero potential depth. The intensity ramps take \SI{130}{\micro\second} each, and the movement is performed in \SI{310}{\micro\second}. During the transport process, the molasses lasers are applied with the same detuning and half the intensity as during the molasses phase for achieving single-atom preparation via light-induced two-body collisions (\cite{Schlosser2002,Brown2019} and references therein). \added{This adds continuous cooling of the atoms, which in general is not necessary for efficient transport \cite{OhldeMello2019}.} After transport operations to every initially empty buffer trap, the atom configuration is recorded using fluorescence imaging as described above. During atom detection, no two-body loss occurs in the buffer traps since there is one atom per buffer trap at most. During cooling, transport, and imaging, in the reservoir, two-body atom loss occurs only at a very low rate, as it is inversely proportional to the trapping volume \cite{Kuppens2000}. Due to the large volume of the reservoir trap, the loss rate is three orders of magnitude smaller than for the traps of the tweezer array.
After imaging, the atom occupation in the buffer traps is further analyzed. For a sufficiently large number of atoms in the reservoir, a probability of 59.6(4)\% for filling the buffer traps with an individual atom is achieved.\\
\indent
In the next functional module of our approach, the individual atoms trapped in the buffer section are used to deterministically fill the empty sites of the target structure. 
To extract an atom from a buffer trap, we direct the transport tweezers to the respective buffer site, ramp the tweezer depth from zero to $U_{\mathrm{T}}/k_{\mathrm{B}} = \SI{1600(400)}{\micro\kelvin}$, move the atom to the intended target site, and release it by ramping the depth to zero again. All intensity ramps and movements have the same timing as for loading of the buffer traps from the reservoir. 
For this atom rearrangement, we use the shortest-move heuristic sorting algorithm described in Ref. \cite{OhldeMello2019}. The transport efficiency between target and buffer traps has been determined to be 75.3(9)\%.
This sequence is repeated until the target structure is completed or the buffer section is empty.
For initiation of a repetitive loading sequence of the target structure, the buffer section is reloaded from the reservoir. From the previous buffer trap occupation and the performed transport moves which empty the buffer traps, we can calculate the resulting atom occupation of the buffer traps. This eliminates the requirement for recording a new image of the atom occupation and thus saves time and avoids atom losses. Using this already available information, reloading of the empty buffer traps is initiated. After this sequence, a fluorescence image such as in Fig.~\ref{fig:Principle} is recorded, updating the complete atom-number status of reservoir, buffer section, and target structure.\\ 
\indent
To demonstrate composition and maintenance of predefined target structures on the underlying grid, we load the six-site hexagon target structure atom by atom, as shown in Fig.~\ref{fig:atom_by_atom}. Following the initial loading of the reservoir (Frame 1), the buffer traps are loaded from the reservoir (Frame 2), showing that, in this realization in four of the seven buffer traps, individual atoms are available for transfer to the target structure. Next, a sequence of six repetitive cycles (Frames 3 to 8) is started in which the target structure is assembled by transferring one atom per cycle from the buffer to the target section. After each cycle, the buffer traps are reloaded, and the next cycle is started by imaging the present trap occupation. \added{One cycle typically takes \SI{230}{\milli\second}, of which about \SI{130}{\milli\second} are necessary for image acquisition and readout. The analysis of the trap occupation and subsequent filling of the target traps typically takes \SI{65}{\milli\second}, while \SI{35}{\milli\second} are necessary for buffer refilling.} For comparison, the lifetime of the atoms in the tweezer array is \SI{10.0(5)}{\second}.\\ 
\indent
For a typical quantum simulation and computation application, one is interested in filling the target structure without defects in the shortest possible time. We have therefore performed a second measurement series, with the goal to fill all target traps from the buffer traps as fast as possible. Evaluating the cumulative success rate for a defect-free target array quantifies in which percentage of realizations the target structure has been achieved in one of the first $n$ rearrangement cycles. For the six-site hexagonal structure, the cumulative success rate is shown in Fig.~\ref{fig:atom_no_evolution_20220225}. After eight rearrangement cycles, we achieve a fully filled structure with 86.8(7)\% probability, while a cumulative success rate of 91.5(6)\% is observed after 15 cycles. Furthermore, Fig.~\ref{fig:atom_no_evolution_20220225} shows the fraction of loaded buffer traps for every rearrangement cycle. As we do not load atoms into the tweezer array directly from the molasses, all buffer and target traps are empty before the first rearrangement cycle. The first filling of the buffer traps gives a filling fraction of 0.596(4), as expected for single-atom preparation via light-assisted collisions for a sufficiently large number of atoms available in the reservoir. For later cycles, the refill probability is influenced by the change in atom number and temperature in the reservoir trap. The atom number is reduced due to generic loss processes occurring in the unperturbed trap \added{[lifetime = \SI{5(1)}{\second}]} but is dominated by atom extraction during the first two to three rearrangement cycles.\\
\indent
From this observation and the fact that the reservoir fluorescence and the buffer-trap filling fraction in Fig.~\ref{fig:atom_no_evolution_20220225} show a strongly correlated temporal behavior, we deduce that limitations of our current setup arise from the relatively small number of atoms in the reservoir trap and the inability of reloading. From the $\sim$80 atoms initially trapped in the reservoir, we prepare on average 10(2) atoms in the buffer section. If continuous or repeated reloading could be established, this would make a constant filling of 60\% of the buffer traps readily available. 
Further improvement of the overall performance could be achieved by increasing the current transport efficiency of 75.3(9)\% between target and buffer traps. Optimizing the experimental conditions should allow for an increase of the transport efficiencies above 98\% as reported in Ref. \cite{Schymik2020}.\\
\indent
To summarize, in this letter, we have presented a method for loading individual atoms from an external reservoir into a tweezer array. We achieve this by repeatedly extracting small atom ensembles from a large dipole trap acting as an atom reservoir using a movable laser beam acting as transport tweezers. During transport, we prepare individual atoms from these ensembles by collisional blockade and store them in the dedicated buffer trap section of our tweezer array. Atoms from this buffer section are transported again to deterministically build up a target structure or refill empty target-structure sites. The reservoir and buffer trap technique can be implemented as an experimental addition to large tweezer arrays which are preloaded with individual atoms by other methods to compensate for atom loss or revoke intentional atom removal.\\
\indent
A straightforward extension of this letter is accessible by the implementation of parallelized operation of the functional modules introduced above. Continuous loading of a tweezer array can be achieved by including a method for repeated refilling of the reservoir trap with cold atoms. Recently, long-distance transport and continuous supply of ultracold atom clouds have been demonstrated in Refs. \cite{Klostermann2022, Chen2022}, matching our requirements.
We envision establishing MOT operation in a separated part of the vacuum system and continuously transporting cold-atom clouds into the reservoir trap by a cold-atom beam line. With this, the reduction of the buffer-section reloading probability due to a decrease in the number of atoms in the reservoir will be eliminated, and the continuous provision of individual atoms to fill the target structure will be maintained at a high rate. 
Combining several cold-atom beam lines delivering different atomic or molecular species with reservoir and tweezer traps holding all species will allow for the efficient implementation of multispecies tweezer arrays for various applications. 

\begin{acknowledgments}
We acknowledge financial support by the Federal Ministry of Education and Research (Grant~No.~13N15981), by the Deutsche Forschungsgemeinschaft [Grants~No.~BI~647/6-1 and No.~BI~647/6-2, Priority Program SPP 1929 (GiRyd)], and by the Open Access Publishing Fund of Technische Universit\"at Darmstadt. We thank the labscript suite \cite{Starkey2013} community for support in implementing state-of-the-art control software for our experiments.
\end{acknowledgments}
\bibliography{Pause_Reservoir}
\end{document}